# Bounded frequency lattices in integrated lithium niobate coupled ring cavities

Hiep X. Dinh[1,2], Armandas Balčytis[1,2*], Guanghui Ren[1,2], Mei Xian Low[1,2], Arnan Mitchell[1,2], and Thach G. Nguyen[1,2]

[1]*Integrated Photonics and Applications Centre, RMIT University, Melbourne, VIC 3000, Australia*
[2]*ARC Centre of Excellence in Optical Microcombs for Breakthrough Science (COMBS), Australia*
armandas.balcytis@rmit.edu.au

**Abstract:**

Synthetic dimensions provide a powerful tool that uses comparatively simple structures to probe high-dimensional topological physics, in which edge states emerging at lattice boundaries are of great importance. However, the demonstration of lattice boundaries in synthetic dimensions is relatively nascent. In this work, we realize an integrated coupled ring system in a thin-film lithium niobate photonic platform that enables the simulation of one-dimensional frequency crystal lattice with sharp boundaries, attaining suppression for two coupling terms with a single auxiliary cavity. Their effect on tight-binding lattice dynamics was verified by acquiring discretized band structures of an $N = 7$ site lattice. The ability to create robust frequency-space boundaries is a key step toward the realization of topological systems that harness bulk-edge correspondence as well as optical information processing in a photonic chip.

## I. Introduction

The synthetic dimension concept has attracted significant efforts for exploring complex topological physical phenomena that pose challenges for probing with real-space systems [1]. It allows the use of low-dimensional devices to simulate high-dimensional physical phenomena by harnessing additional degrees of freedom [2, 3, 4, 5] as a supplement to spatial dimensions. In photonics, properties of light, such as frequency [6, 7], time bin [8], polarization [9], spin states [10], orbital angular momentum [11] and waveguide mode [12] have been harnessed for constructing independent photonic state connectivity pathways.

A widely-used method for demonstrating lattice dynamics in a photonic frequency space is to use ring resonators incorporating electro-optic modulation [10, 13, 14, 15, 16]. By dynamically modulating the resonator cavity, one can create reconfigurable frequency crystal lattices with versatile coupling range and strength as well as tunable gauge potentials, enabling the simulation of complex structures [17, 18, 19, 20, 21]. It is important to note that in such a device the synthetic lattice can extend towards either ascending or descending ring mode frequencies without constraint. In topology science, however, boundaries are fundamental in accessing bulk-edge correspondence [5]. Edge states and edge effects that exist at boundaries, are also of practical interest [22], with potential applications such as topological lasing, photonic delay lines [23] and optical isolation [7, 24]. However, the ubiquitous approach for establishing a synthetic frequency dimension using a modulated ring cavity does not inherently exhibit a boundary in a way that a real-space multi-ring counterpart device would [25].

Synthetic boundaries in photonic frequency dimensions can be realized by engineering the mode structure of a resonant cavity to produce a selective mismatch between mode spacing and driving RF frequencies. Notable demonstrations of the concept in an integrated photonic platform include ring cavities in which equidistant mode spacing is disrupted via splitting due to coupling between degenerate traverse-magnetic and traverse-electrical modes [26], or between clockwise and counterclockwise waves induced by a photonic crystal section [27]. Their drawback is that the resultant boundary is sensitive to fabrication imperfections and has scant options for control and reconfigurability. Another promising approach leverages coupled rings, early realizations of which were achieved using optical fiber cavities [28] or an integrated photonic platform [26]. The optical fiber implementations offer experimental flexibility but are susceptible to loss of coherence due to environmental interference over macroscale cavity lengths commonly extending over tens of meters. Taking the next step in integrating the entire synthetic frequency dimension device on an electro-optic photonic waveguide platform, such as thin-film lithium niobate on insulator (LNOI), enables the leveraging of the full range of its properties - including broad wavelength spectrum, ultra-low waveguide propagation loss, strong nonlinear response, and high power handling capability - making it a default choice for demanding electro-optic applications [29].

We report on an LNOI chip integrated device comprised of two coupled asymmetric ring cavities and use it to demonstrate robust frequency dimension boundaries that delimit finite resonator mode ladder segments. By applying a sinusoidal modulation signal oscillating at a 9.2 GHz rate, matching the frequency spacing of the primary ring, to the integrated phase modulator section of the coupled ring device, the establishment of a bounded one-dimensional synthetic tight-binding lattice in the frequency dimension is observed for both nearest and next-nearest mode coupling. This is achieved by leveraging a combination of ring mode splitting to induce a local mode displacement as well as coupling-induced dispersion to attain a global shift in cavity resonances. Effectiveness of frequency space boundaries is confirmed by measuring steady-state spectral response to pumping at various lattice sites and through the corresponding band structures obtained by detecting the time-resolved ring transmittance under dynamic modulation.

## II. Device concept and design

A ring resonator equipped with an integrated phase modulator section possesses a discrete set of modes with a uniform spectral spacing that can be reinterpreted as tight-binding model lattice sites in a frequency dimension photonic analog to real-space coordinates [13, 17, 19]. Photon hopping between adjacent sites in a lattice is induced by some temporal perturbation, such as applying a radio-frequency (RF) modulation signal that matches an integral multiple of the ring free-spectral range (FSR). With continuous driving, movement of photons across ring modes occurs symmetrically in both directions, mimicking particle random walk, as illustrated in Figure 1(a). Constraints for frequency space photon propagation result primarily from device imperfections – photon decay through scattering or absorption loss in the optical waveguide as well as a drop-off in coupling efficiency caused by the detuning between a fixed modulation frequency and a frequency dependent ring FSR dispersion [26]. However, neither of these effects is sufficiently abrupt to act as a lattice boundary.

An effective synthetic frequency dimension boundary requires a sharp discontinuity in the ring resonator mode structure to disrupt its translation symmetry along the photon energy axis. This can be achieved by introducing a localized lattice defect to suppress RF mediated inter-mode coupling and halt further photon frequency shifts. Integrated ring device implementations can leverage a variety of dispersion control mechanisms, such as the mode splitting phenomenon [26, 28]. As illustrated in Figure 1(b), it can occur when a mode in the main ring A couples to a frequency-matched mode in auxiliary ring B and hybridizes. This creates a pair of different symmetry modes with opposite frequency shifts relative to the initial value that scales with the coupling strength between the two rings. The ring mode structure will locally deviate from the uniform spacing required for continuous RF-mediated photon hopping, and, in the absence of a resonant mode at the expected FSR increment, the photon hopping rate from an adjacent mode to the defect site will be significantly lower than in the rest of the lattice. As inter-site hopping in the opposite direction from the split-mode defect remains unimpeded, a frequency space analog to reflection from a boundary can be expected.

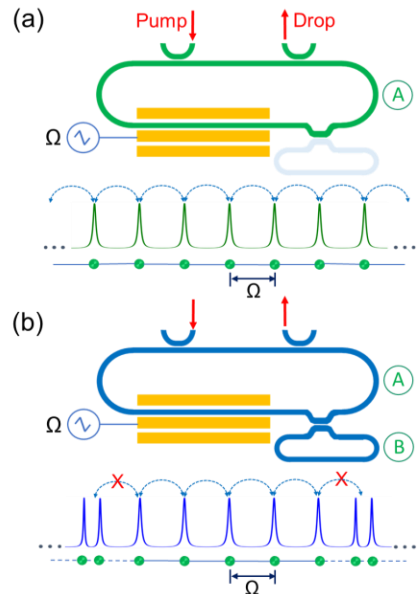

*Figure 1 Illustration of frequency-space lattices and their boundaries in sketched modulated ring cavities and their transmittance spectra. (a) Unbounded case with a single electro-optically modulated ring producing a 1D tight-binding lattice with uniform coupling strength between sites. (b) Asymmetrical coupled ring cavities where mode frequency spacing continuity is periodically disrupted by mode splitting, establishing a bounded lattice segment.*

Implementation of the frequency lattice boundary concept in an X-cut thin-film LNOI waveguide platform is shown in Figure 2(a). The device consists of two micro-resonators of different circumferences. The total length of the integrated auxiliary-main ring cavity system is approximately $l$ = 15.9 mm. This represents a three-order-of-magnitude reduction in optical path length over the pioneering optical fiber loop realization of 38.6 m [28]. Fabrication of the integrated circuit is based on the strip-loaded composite waveguide approach [30, 31]. The material stack used in this process is illustrated in Figure 2(b), where optical mode confinement is achieved by lithographically defining a 300 nm-thick and 1 μm-wide silicon nitride ridge on a 300 nm-thick $LiNbO_3$ layer, shown in right panel of Figure 2(c). Strip-loading offers advantages such as a comparatively simpler fabrication process and vertical waveguide sidewalls, conducive to reproducible directional coupler performance, however, at the cost of somewhat lower electro-optic modulation efficiency due to lower mode confinement in the $LiNbO_3$ layer, comparing to etched $LiNbO_3$ waveguide configuration.

The primary ring A innately has a $FSR_A$ = 9.5 GHz spaced set of equidistant frequency modes comprising the synthetic frequency crystal lattice. To facilitate inter-site hopping, the primary ring comprises a broadband phase modulator encompassing one of the racetrack cavity straight sections. The phase modulator is arranged as an approximately 6.5 mm-length GSG co-planar transmission line along the Y-axis of $LiNbO_3$ crystal. Electrodes were formed by lithographically patterning a 300 nm thick gold layer. Their 7-μm spacing gap, depicted in the leftmost panel in Figure 2(c), was chosen to minimize waveguide propagation loss induced by metal absorption while maintaining a low modulator switching voltage.

Mode structure of the primary ring A was engineered by coupling it to an auxiliary ring B with a larger $FSR_B$ = 76 GHz. This value is 8 times that of main ring, so that a recurrent mode splitting interaction, hence a frequency space boundary, is produced with a $FSR_B = 8 \times FSR_A$ repetition. As a result, the primary ring resonance spectrum is partitioned into segments of seven lattice sites, terminated by boundaries on each end. This $FSR_A$ and $FSR_B$ ratio was chosen to ensure feasibility of time-resolved tight-binding lattice band structure acquisition, outlined in Section IV, which are subject to distortions for longer lattice segments due to detector bandwidth limitations. The coupling strength between the primary and auxiliary rings directly governs the mode splitting magnitude, which, to enforce an effective boundary, should be well above the full-width half-maximum of a typical main cavity mode. Furthermore, strong coupling related dispersion causes slight shifts in mode frequencies beyond the boundary so that the driving RF frequency becomes incommensurate with their spacing outside of the bounded section. The power coupling coefficient between two rings is controlled by the gap and interaction length of the directional coupler and set in the present case to $\kappa = 0.6$. Resonances of the auxiliary ring were tuned in to match those in the main ring using a thermal heater, shown in Figure 2(c).

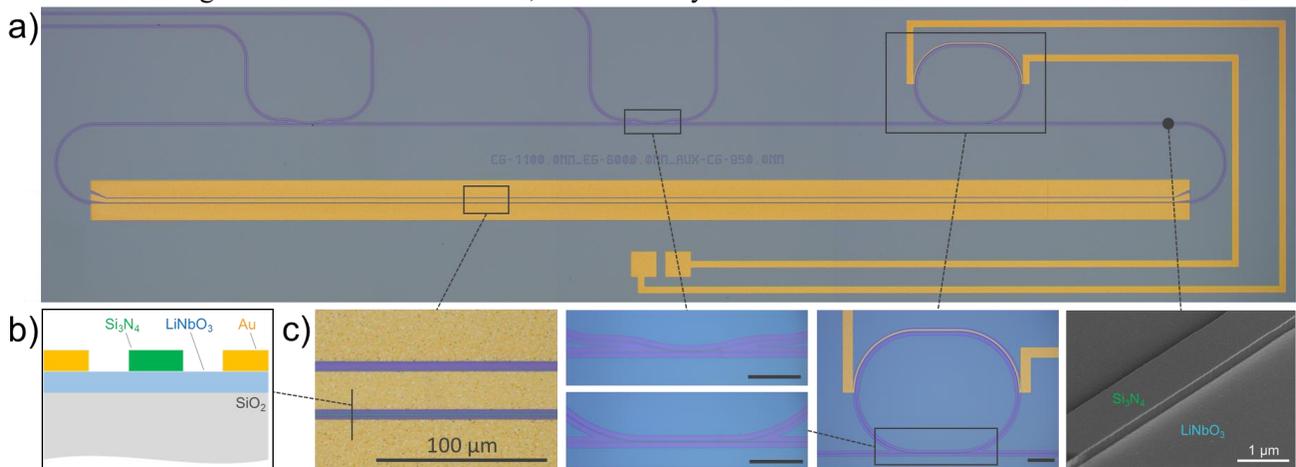

*Figure 2 Integrated LNOI ring resonator device implementing frequency dimension lattice boundaries.* (a) Micrograph image of coupled asymmetric cavity ring resonator device with a traveling wave RF electrode. (b) Sketch of SiN loaded LNOI waveguide modulator section illustrating the material stack. (c) Magnified images of the highlighted sections (from left-to-right) - phase modulator electrodes, directional couplers interfacing bus waveguides and main-auxiliary rings, auxiliary ring with heater for thermal tuning, and SEM image of the SiN strip-loaded LNOI waveguide.

### III. Experimental characterization.

The mode structure of a ring cavity device, from which a synthetic frequency space lattice is to be constructed, is revealed by probing its optical transmittance. Such measurements in the telecommunication C-band region were performed via the cavity drop port for the single and coupled ring devices sketched in Figure 1, and are plotted in Figure 3(a). In general terms, the cavity exhibits an extinction ratio of approximately 20 dB, and a loaded quality factor of approximately $Q = 3.1 \cdot 10^5$, corresponding to a photon decay rate of $2\gamma = 0.7$ GHz.

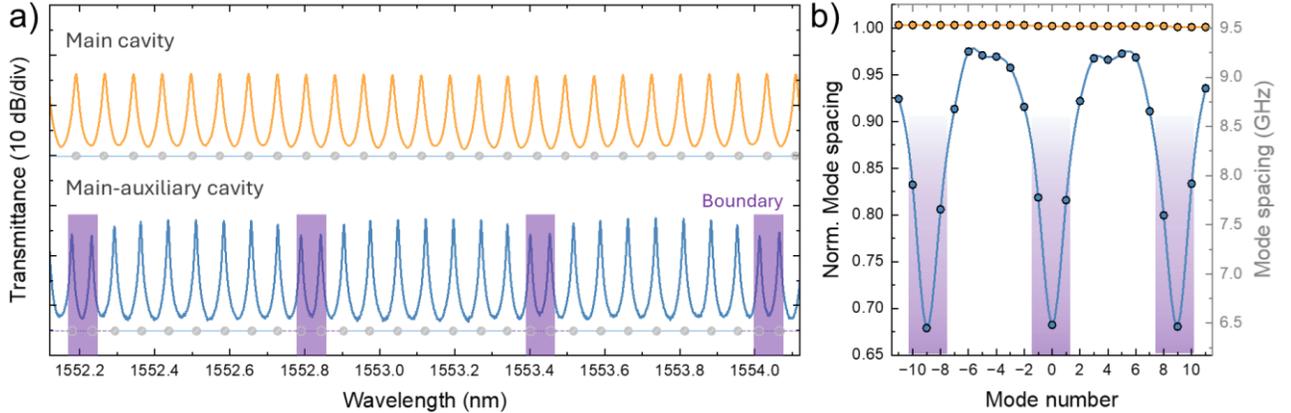

*Figure 3 Ring cavity optical characterization. (a) Comparison of drop-port optical transmittance of single and composite ring devices in the 1550 nm wavelength range. Highlighted spectral segments show auxiliary ring coupling induced mode splitting. (b) Variation in ring mode spacing due to dispersion induced by auxiliary ring coupling. Single ring mode spacing was used as a normalization reference.*

Contrasting the transmittance of a single ring cavity to its composite counterpart with a coupled auxiliary ring, shown in Figure 3(a), it can be noted that a single ring possesses a set of resonances with a uniform arrangement throughout the probed spectral region. Such a spectrum represents the typical 1D tight-binding frequency lattice arrangement, where a single FSR-matched coupling mechanism can link the full range of modes. The sketched lattice chain illustrates this uniform connectivity regime. Conversely, the composite cavity, in which the primary and auxiliary ring cavity resonances are made to line up every 8×$FSR_A$ exhibits hybridization-induced recurrently split modes that interrupt continuous resonance spacing and isolate the lattice chain into seven-site segments. The inter-ring coupling strength is estimated at $\mu = 3.4$ GHz.

As indicated by the mode frequency spacing values over a 20×$FSR_A$ range in Figure 3(b), local mode hybridization induces three related modifications of the main cavity spectrum. First of these is the direct splitting of the affected mode, resulting in the lowest adjacent mode spacing of 0.67, normalized to the nominal 9.5 GHz $FSR_A$. These local regions with a smaller mode spacing cause an abrupt detuning with an RF signal. In addition, primary cavity mode structure is affected even in the absence of direct hybridization, showing small deviations in spacing by a factor between 0.92 and 0.97, with modes closest to the boundary affected most. This can be attributed to the auxiliary ring coupling-related phase shifts, which can extend over a broader spectral range than changes in resonance frequency or group velocity [33]. In contrast, mode spacing in a single ring accumulates changes only gradually due to small waveguide group velocity dispersion, so that a nearly constant $FSR_A$ is maintained.

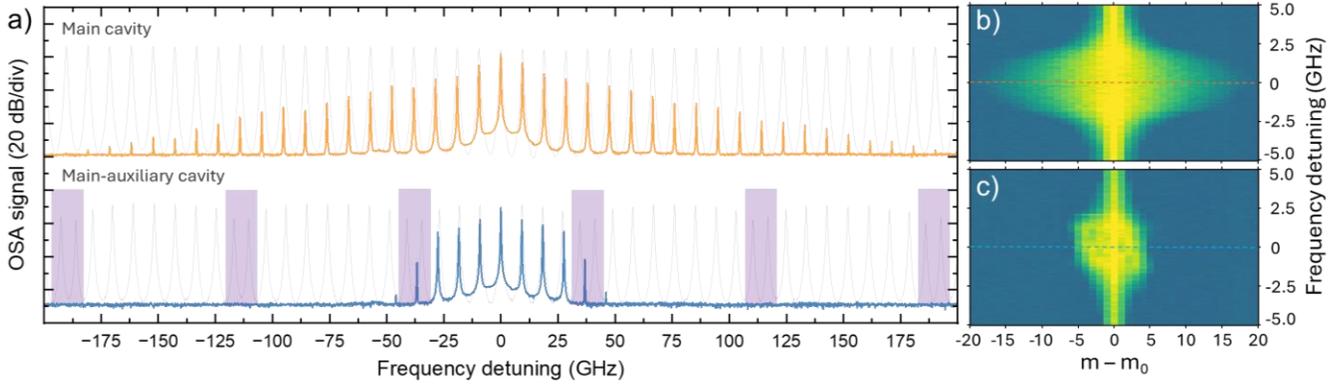

*Figure 4 Electro-optical response of modulated ring cavities. (a) Comparison of spectral outputs of driven ring devices with on-resonance laser pumping. Orange line represents a single unperturbed ring, whereas blue line shows the main-auxiliary composite cavity output. Transmittance of the respective devices, along with location of boundaries, are outlined in the background as a guide for the eye and are not-to-scale. Frequency dimension lattice site amplitude plots, showing intensity of the different ring modes and their changes with pump laser detuning for (b) single main modulated cavity, and (c) main-auxiliary coupled cavity.*

A terminated frequency space tight-binding lattice can be demonstrated by phase-modulating the cavity with a sinusoidal signal of the form $V(t) = V\cos(\Omega t)$ [17, 19]. Here the frequency was set to $\Omega \approx 9.2$ GHz, to match the auxiliary ring coupling modified resonator mode spacing. This resonance condition supports efficient photon hopping between adjacent frequency modes, effectively emulating lattice nearest-neighbor coupling in a photonic synthetic frequency dimension.

The steady-state electro-optic response spectra of the coupled primary-auxiliary, as well as a single ring for reference were measured using an optical spectrum analyzer under moderate 11 dBm optical pumping powers and driving with a 20 dBm RF power of FSR-matched modulating signals. Figure 4(a) shows the resulting spectra of the coupled main-auxiliary ring (blue) and of the single main ring (orange). A single ring exhibits up to 40 sidebands that conform to continuous random walk hopping transitions away from the pumping site, without boundary constraints. The 2.3 dB-per-line decrease in sideband power across the spectrum is primarily caused by waveguide propagation loss, rather than group velocity dispersion induced detuning [34].

In contrast, the main-auxiliary coupled ring cavity exhibits a constrained spectrum with a limited $N = 7$ number of sidebands. Within the bounded spectrum, intensity difference between adjacent sidebands is 2.3 dB, the same as in the case of the single ring. Outside the boundaries, however, local and global mismatch of RF drive signal frequency to resonant mode spacing inhibits frequency conversion. The formation of abrupt frequency lattice boundaries is verified by a sudden drop in comb line power, with an extinction level of approximately 16 dB.

A comprehensive picture of the terminated frequency dimension lattice can be visualized by 2D plots of optical spectrum analyzer-derived steady-state mode peak intensities, and their variance when the pump laser is scanned across a single FSR range. For the single ring case in Figure 4(b), without auxiliary ring induced mode engineering, the output exhibits a significant number of populated modes. Their amplitude tends to decay exponentially away from the pump depending on the ratio between RF inter-mode coupling strength $J$ and the propagation loss rate $\gamma$. Conversely, mode intensity variation across the vertical optical pump detuning axis is determined by differences in optical power injected into the cavity due to resonant ring filtering.

The optical heterodyne spectral plot measured on the main-auxiliary coupled ring device, shown in Figure 4(c), indicates that the frequency random walk is confined in an $N = 7$ photonic lattice region and is largely unaffected by changes in pump laser frequency. There are also indications of interference fringes in mode intensity, arising due to interactions between photons reflected at the boundaries.

Frequency dimension boundary robustness is verified by measuring the steady-state spectra of the coupled ring system when the optical pump is aligned to modes adjacent to the boundaries, as shown in Figure 5(a). While photon hopping towards the lattice bulk proceeds with the 2.3 dB-per-line exponential decay rate, frequency conversion towards the boundary produces a similar 16 dB reduction in sideband intensity, to that observed when pumping a central lattice site. Using the frequency mirror characterization model [26], the reflectivity of the boundary in our device was estimated at $R = 0.9988$. The steady-state spectra when optical pumping is performed at other lattice site modes are outlined in Supplementary Figures 2-3.

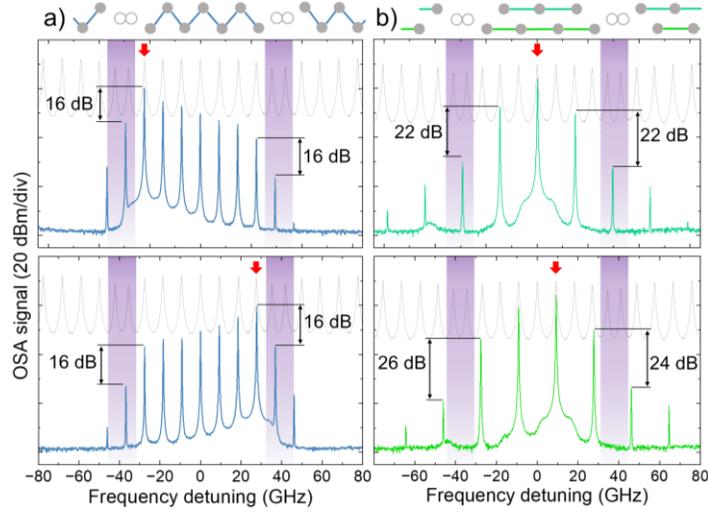

*Figure 5 **Electro-optical response of modulated dual ring cavities different pumping conditions.** Output of the ring system modulated at 20dBm on-chip RF power and 10dBm on-resonant optical pumping at different modes. (a) Nearest-site 1×FSR driven case in which lattice is pumped at opposing endpoints. (b) Next-nearest-site 2×FSR modulation when input pump is tuned to each of the two sublattices. Top sketch outlines lattice partitioning schemes. Gray line outlines the qualitative transmittance spectrum. Red arrows indicate the pumped modes.*

Of note is that the auxiliary cavity-induced boundary is effective for next-nearest lattice site coupling, enacted by a $2\Omega \approx 18.4$ GHz driving. As sketched in Figure 5(b), 2×FSR modulation splits the $N = 7$ lattice into two $N' = 3$ and $N'' = 4$ length sublattices. Each can be pumped separately and exhibits the corresponding number of spectral peaks, with a sharp spectral intensity drop observed beyond the bounded region, irrespective of which specific mode is pumped. Conversely, the dual ring cavity device was ineffective in suppressing 3×FSR coupling, as $3\Omega \approx 27.6$ GHz coincided with the boundary spectral span. We expect this match can be avoided through a combination of higher cavity finesse and tailoring cavity coupling strength to ensure mode splitting produces a boundary with a span incommensurate with relevant $n$×FSR modulation signals.

## IV. Synthetic-space band structure spectroscopy

The range of permissible energy states in a translationally symmetric lattice is expressed as one or more bands mapped to a Bloch quasi-momentum $k$ inverse space. For a modulated ring cavity synthetic frequency dimension device, this wavenumber maps directly to the time variable [13]. Hamiltonian for the one-dimensional tight-binding lattice, is given by the general expression [28]:

$$H = \sum_N \varepsilon \hat{b}_m^\dagger \hat{b}_m - \sum_N J_\eta (\hat{b}_m^\dagger \hat{b}_{m+\eta} + \hat{b}_{m+\eta}^\dagger \hat{b}_m), \tag{1}$$

where $b_m$ is an operator for particle creation at site $m$, $\varepsilon$ is the on-site potential, and $J_\eta$ is the coupling rate between lattice sites spaced by $\eta$×FSR in frequency. The total number of lattice sites are represented by $N$, so that both infinite $N \to \infty$ and bounded cases are modelled. The experimental implementation in this work operates with nearest $J_1$ and next-nearest-neighbor $J_2$ coupling, and the length of the bounded 1D lattice is $N = 7$.

Synthetic dimension band structure spectroscopy was performed by measuring the time-resolved transmittance of the modulated ring cavity system [13]. This was done by repeatedly measuring transmitted intensity using a fast 75 GHz bandwidth photodetector as the pump laser wavelength was swept incrementally. Details on the measurement setup are provided in Supplementary Figure 1. Figure 6 shows examples of the resulting synthetic band structure plots for the quasi-infinite and bounded $N = 7$ lattices in the *1*×FSR and *2*×FSR coupling range cases and compares them with predictions simulated using the steady-state matrix approach [35]. The reciprocal space Brillouin zone is defined as a single optical round-trip period in the time-resolved transmittance. Here, composite primary-auxiliary ring devices were driven by 25 dBm 9.2 GHz and 18 dBm 18.4 GHz RF signals, establishing respective frequency space coupling strengths of $J_1 \approx 0.6$ GHz and $J_2 \approx 0.4$ GHz.

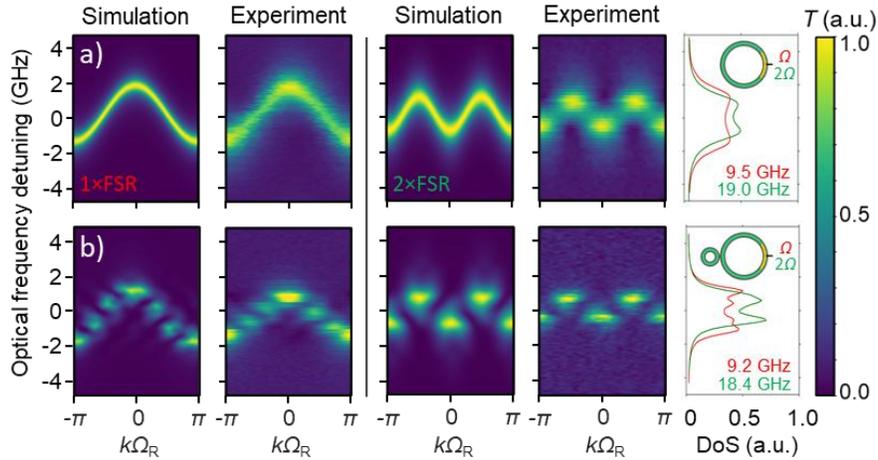

*Figure 6 Band structure measurements of synthetic frequency dimension lattices in LNOI ring cavities. (a) Unterminated 1D tight-binding lattice in a single ring. (b) Finite N = 7 site 1D lattice with frequency space boundaries created by auxiliary ring couplings. Each panel from left to right shows simulated and measured band structures for 1×FSR and 2×FSR coupling cases, as well as the respective density of states for each device, sketched in insets.*

Figure 6(a) illustrates the reference case of a single modulated ring without any auxiliary cavity coupling-induced mode splitting effects. Continuous tight-binding lattice band structures [19], with no clear distinction between individual eigenstates, are observed. In contrast, Figure 6(b) gives the band structure of a $N = 7$ finite lattice with a discretized set of energy states, positioned on the sinusoidal tight-binding lattice dispersion curve envelope. The number of these states, seven and three for the respective coupling ranges, matches that of populated sites in the bounded lattice. Curves on the right of Figure 6 show a representation of the density-of-states of the two ring devices, obtained as a time-averaged transmittance measurement of the modulated rings. Here, the bounded lattice reveals a fine structure absent for the single ring device, further indicating discrete energy states.

However, these discrete energy states exhibit a degree of dispersion along the quasi-momentum axis. This can be attributed to mode spacing variance the main-auxiliary ring system in the vicinity of the frequency space boundary, shown in Figure 3(b). As a result of a local FSR and RF driving frequency mismatch, band structure measurements are dynamic [36] due to a strong relative contribution of the affected edge modes in a short $N = 7$ lattice. As revealed by the broader time-resolved transmittance scan, shown in Supplementary Figure 4, arrangement of the energy states in the band structure depends on proximity of the probed mode to the lattice boundary. Furthermore, the rate of their dynamic oscillations increases with the associated FSR detuning. All experimental observations are in notable agreement with the simulated band structures accounting for these effects.

Variations in tight-binding lattice nearest-neighbor coupling strength $J$ in a frequency dimension device can be tuned by adjusting the driving RF signal power, as shown in Figure 7. The optical frequency comb spectra corresponding to each of the band structures in Figure 7(a) are provided in Supplementary Figure 3. The flat band when the modulating signal is switched off gradually attains the sinusoidal tight-binding model dispersion curve shape as RF power is incremented. With a driving RF signal >10 dBm, the frequency lattice photon hopping rate $J \approx 0.2$ GHz becomes sufficient to overcome propagation loss and reach lattice boundaries. This coincides with the emergence of observable discrete states, which become more defined as RF power is ramped up further. The strong coupling between primary and auxiliary rings produces a large mode splitting, ensuring $J < \mu$ even at maximum attainable driving power. Hence, no loss of confinement in the bounded region has been observed.

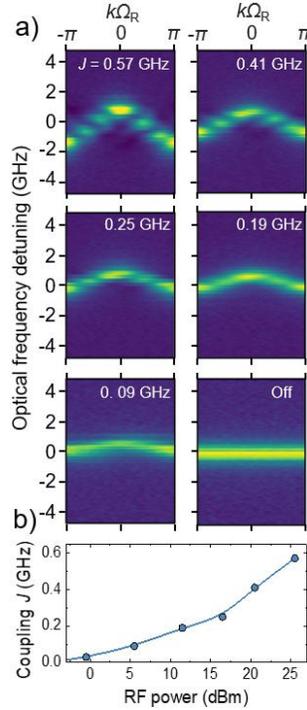

*Figure 7 Band structures of a finite frequency dimension lattice at different driving RF powers. (a) Experimental band structures corresponding to a range of mode coupling strengths. (b) Relationship between driving on-chip RF signal power and the coupling strength J estimating via tight-binding model fitting.*

## VI. Discussion

We designed and fabricated a coupled asymmetric micro-ring resonator cavity system in an integrated LNOI platform, and used it to experimentally demonstrate a simulation of frequency space boundaries in a synthetic 1D tight-binding model lattice. Such modulated ring resonator devices on the LNOI platform represent promising building blocks for higher complexity integrated circuits for both fundamental modelling of topological and quantum systems, as well as for uses in electro-optic frequency comb spectral engineering, optical isolation [7] as well as photonic computation, such as arbitrary linear transformations [37], convolution processing [38], and steered quantum walks.

One advantage of the auxiliary ring-induced mode splitting approach to frequency space boundary formation lies in its flexibility. For instance, auxiliary ring $FSR_B$ can be chosen as non-commensurable to that of the main ring $FSR_A$, so that only a single defect mode frequency mirror is created. Conversely, matched $FSR_B = n \times FSR_A$ combinations can be employed either to create bounded lattices or to engineer alternating mode spacings [39]. Furthermore, on-chip EO or thermal tuning provides control over ring FSR and coupling strength, which can be leveraged to realize adjustable or switchable frequency mirrors. In addition, our device implementation enabled a boundary for both nearest $J_1$ and next-nearest $J_2$ coupling terms. This work suggests that effectively suppressing longer range coupling terms could be possible with a single auxiliary cavity. Precise mode engineering that leverages a combination of mode splitting and waveguide group velocity and mode coupling-induced dispersion effects can achieve an overall broader boundary without the added complexity of purely employing mode splitting with multiple rings.

Further development directions of on-chip synthetic frequency dimension boundaries include better control over mode coupling induced dispersion, particularly its influence on lattice sites in the vicinity of the boundary. This variance in FSR values scales mainly with auxiliary ring coupling strength, therefore, can be minimized by optimizing cavity interaction. Furthermore, by utilizing the variety of dispersion engineering options present in integrated platforms with strong optical confinement, it is feasible to ensure spectrally localized mode splitting and dispersion. Realizing frequency boundaries that remain effective for larger $n \times FSR$ coupling ranges is an important step towards demonstrating higher-dimensional topological models and more efficient optical processing capabilities.

## Acknowledgements

This work was performed in part at the Melbourne Centre for Nanofabrication (MCN) in the Victorian Node of the Australian National Fabrication Facility (ANFF). This research was conducted in part by the Australian Research Council Centre of Excellence in Optical Microcombs for Breakthrough Science (project number CE230100006) and funded by the Australian Government. The authors would like to thank Prof. Tomoki Ozawa, Prof. Satoshi Iwamoto, A/Prof. Yasutomo Ota, and Prof. Toshihiko Baba for valuable discussions.

**Additional information:** Supplementary Information accompanies this paper at xxx.

**Competing interests:** The authors declare no competing interests.

**Data availability:** The data that support the findings of this study are available from the corresponding author upon reasonable request.